# Protocol for a Study of the Effect of Surface Mining in Central Appalachia on Adverse Birth Outcomes

Dylan S. Small[1]
Dan Firth
Luke Keele
Matthew Huber
Molly Passarella
Scott Lorch
Heather Burris

Abstract

Surface mining has become a major method of coal mining in Central Appalachia alongside the traditional underground mining. Concerns have been raised about the health effects of this surface mining, particularly mountaintop removal mining where coal is mined upon steep mountaintops by removing the mountaintop through clearcutting forests and explosives. We have designed a matched observational study to assess the effects of surface mining in Central Appalachia on adverse birth outcomes. This protocol describes for the study the background and motivation, the sample selection and the analysis plan.

---

[1] Address for correspondence: Department of Statistics, The Wharton School, 400 Huntsman Hall, 3730 Walnut St., Philadelphia PA 19104. E-mail: dsmall@wharton.upenn.edu

*Background and Motivation for Study*:

Surface mining is a major method of coal mining in Central Appalachia alongside the traditional underground mining because it is typically faster, cheaper and less labor intensive where coal seams are near the surface (Holtzman, 2011). Large scale surface mining in Central Appalachia includes mountaintop removal mining (MRM) where coal is mined upon steep mountaintops by removing the mountaintop through clearcutting forests and explosives, and then using surface mining techniques once the coal is exposed (Pericak et al., 2018). The removed overburden from the mountaintop is deposited in nearby valleys where it buries existing streams. There is evidence for negative environmental effects of large scale surface mining, including MRM, on water quality and air quality (Palmer et al., 2010), and concerns about its effect on human health have been raised. Other types of surface mining in Central Appalachia include area, steep slope, contour, highwall and auger mining, all of which may contribute to negative impacts on human health due to factors such as dust emission.

Hendryx and collaborators have published a series of studies presenting evidence for negative effects of MRM on human health, e.g., Esch and Hendryx (2011) and Hendryx and Holland (2016). Other authors have not found negative effects, e.g., Borak et al. (2012).

A systematic review found it "could not reach conclusions on community health effects of MRM because of the strong potential for bias in the current body of human literature." (Boyles et al., 2017). In particular, the systematic review raised concerns about existing studies with respect to the following potential sources of bias:

(1) Exposures were characterized indirectly and were inconsistently matched to the time-frame for the outcome assessment;
(2) There was a lack of accounting for confounding or modifying variables;
(3) There was a lack of blinding for self-reported outcomes
(4) There was a potential influence of funding source on the authors' interpretation or results and conclusions.

We plan to conduct an observational study of the effect of surface mining in Central Appalachia, (which includes MRM) on low birth weight and other birth outcomes that will mitigate the concerns regarding some of these biases. While our exposure characterization remains indirect – county level yearly surface mining activity -- we will use two exposure characterizations ( county-level surface mining coal production and a geographic information systems (GIS) method) to estimate community proximity to surface coal mining (Pericak et al., 2018). Our exposure assessment is tightly tied to the time frame for outcome assessment since we have yearly exposure data and we know that the time frame in which the exposure matters for the outcome is the approximately nine months of pregnancy before the birth. We control for individual level confounding variables (maternal race and age as well as multiple gestations and infant sex). Also we match high surface mining counties to low/no surface mining counties on county-level confounding variables (poverty rate, median income, percent white, county level education and maternal smoking rate). We also use a difference-in-difference design to control for unmeasured time-invariant confounding. Our primary outcome of birth weight is objectively measured and not self-reported. To minimize the effect of our own views on the results, we are conducting the observational study in a blinded way as recommended by Rubin (2007). We are posting our protocol on ArXiv after matching high surface mining counties to low/no surface mining counties but before examining the outcome data.

*Treated and Control Group, Treated and Control Periods*
Following Hendyrx and Holland (2016), our study population consists of counties in the four states where MRM has been practiced – Kentucky, Tennessee, Virginia and West Virginia.

For our primary analysis, we will consider as treated (high surface mining counties in which MRM is practiced) counties those counties in Central Appalachia coal fields whose average disturbed area percentage from 1999-2011 is over 1% and whose average disturbed area percentage from 1999-2011 is greater than that of 1985-1989 (all counties who met the first criteria met the second criteria). We chose 1999-2011 to be the treated period based on the below graph of the average disturbed area fraction of the 67 counties who had some disturbed area between 1995-2018.

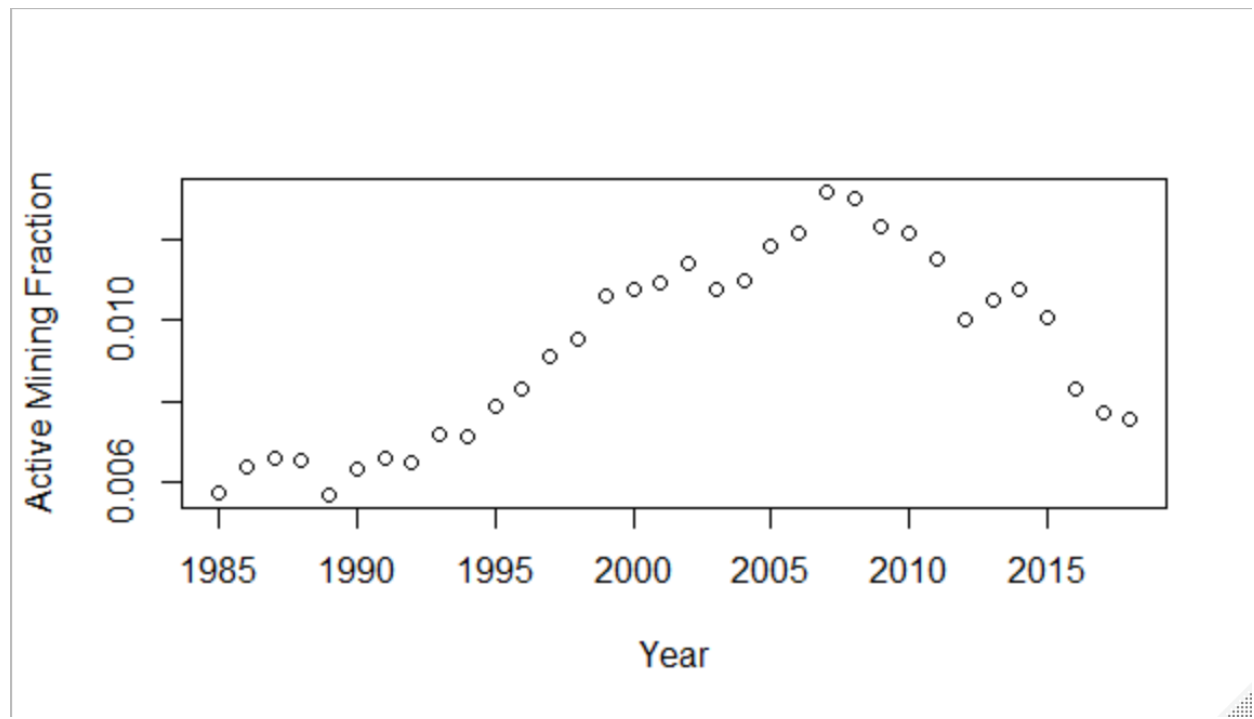

For our control period, we will use 1977-1989. This is motivated by Hendryx and Holland (2016) who used the Clean Air Act Amendment to demarcate a pre and post period. Amendments to the Clean Air Act Amendment were implemented in 1990 in the U.S. with the intent to reduce acid rain and other pollution consequences of burning coal in power plants. Coal reserves in the Central Appalachian region are relatively low in sulfur content and became more financially attractive after these amendments took effect (Hendryx and Holland, 2016). In the later 1990s, MRM also increased because of the development of bigger equipment that enabled mining at scale (Bozorgebrahimi et al., 2003; Ramani, 2012). We use 1999-2011 as the treated period as discussed above because it is a period of high and sustained surface mining. We use 1977-1989 as the control period because it is a period before the Clean Air Act Amendment of the same length as the treated period and a period in which surface mining was not particularly increasing. Evidence for surface mining not particularly increasing in this period comes from the above figure for 1985-1989 and the finding that the acres directly impacted by MRM in Central Appalachia only increased by 9% from 1976 to 1985 as compared to increasing by 159% from 1985 to 1995 and 36% from 1995 to 2005, while the acres directly impacted by other surface mining decreased by 18% from 1976 to 1985 as compared to increasing by 1% and 4% from 1985 to 1995 and 1995 to 2005 respectively (Amos, 2009). In secondary analyses, we will consider treated periods of 1990-1998 and 2012-2017.

These are the 23 treated counties ordered by average disturbed area from 1999-2011.

```
                 county avg.1999.2011 avg.1985.1989
1       Boone County, WV         0.068         0.019
2       Perry County, KY         0.067         0.039
3      Martin County, KY         0.061         0.037
4       Knott County, KY         0.056         0.025
5       Logan County, WV         0.047         0.016
6       Mingo County, WV         0.044         0.012
7        Pike County, KY         0.043         0.015
8        Wise County, VA         0.039         0.005
9      Leslie County, KY         0.032         0.018
10  Breathitt County, KY         0.029         0.024
11    Letcher County, KY         0.028         0.018
12       Bell County, KY         0.021         0.020
13     Harlan County, KY         0.019         0.011
14   Nicholas County, WV         0.019         0.006
15      Floyd County, KY         0.018         0.015
16    Wyoming County, WV         0.016         0.005
17       Clay County, WV         0.015         0.001
18   McDowell County, WV         0.014         0.006
19    Kanawha County, WV         0.012         0.007
20    Lincoln County, WV         0.012         0.000
21    Raleigh County, WV         0.012         0.003
22    Webster County, WV         0.012         0.001
23    Fayette County, WV         0.010         0.004
```

For control counties, we consider counties with a yearly average of less than 1000 short tons of surface mining production per square mile from 1999-2011 (all of the treated counties except for Lincoln County, West Virginia have greater than 1000 short tons of surface mining production per square mile from 1999-2011) and less than 5000 short tons of total production.[2]

These are counties that were excluded that were in Western KY and Northern WV, and weren't included in our disturbed area data set because they are not in Central Appalachia:

| Muhlenberg Kentucky |
| Henderson Kentucky |
| Ohio Kentucky |
| Hopkins Kentucky |
| Brooke West Virginia |

---

[2] There are 34 counties that have an average of greater than 1000 short tons of surface coal production from 1999-2011 and 34 counties that have an average of greater than 5000 short tons of total coal production from 1999-2011. Total coal production is relevant as some of the harmful health effects of coal could come from transporting coal.

| |
|---|
| Monongalia West Virginia |
| Marshall West Virigina |
| Marion West Virigina |
| Union Kentucky |
| Harrison West Virginia |
| Ohio West Virginia |

These are counties that are in Central Appalachia and included in our disturbed area data set, but did not have an average disturbed area percentage from 1999-2011 of over 1%

  Buchanan Virginia
  Magoffin Kentucky
  Wayne West Virginia
  Claiborne Tennessee
  Lawrence Kentucky
  Dickenson Virginia

We compare high surface mining counties to low/no surface mining counties rather than moderately high surface mining counties to moderately low surface mining counties because making the treated and control groups sharply differ in dose increases the insensitivity of the study to unobserved biases when there is a treatment effect and no bias (i.e., it increases the design sensitivity, Rosenbaum, 2004).

*Covariates to control for*

Variables to match counties on:
(i) Average (Mean) of median household income in 1979 and median household income in 1989
(ii) Average (Mean) of median household income in 1999 and median household income in 2011
(iii) Average (Mean) of poverty rate in 1980 and poverty rate in 1990
(iv) Average (Mean) of poverty rate in 2000 and poverty rate in 2010
(v) Percent white in 1990
(vi) Average (Mean) of average education level in 1980 and education level in 1990 where the average education level is 10*Percent less than high school + 12 * Percent with high school degree only +14 * Percent with some college + 16* Percent college grad. These weights are chosen as equally spaced representative values of the amount of education that a person in the category would have, i.e., a person who is a high school dropout might have 10 years of education, a person who is a high school grad might have 12 years of education, a person who has some college might have 14 years of education and a person who is a college grad might have 16 years of education.
(vii) Average (Mean) of average education level in 2000 and education level in 2012-2016
(viii) Smoking Rate among Pregnant Mothers from 1989-2003[3]

---

[3] Maternal smoking during pregnancy was first recorded on birth certificates in 1989. Starting in 2004, Kentucky and Tennessee moved to using the 2003 revised birth certificate. Virginia moved to the 2003 revised birth certificate in mid-year 2012 and West Virginia moved in 2014. Maternal smoking during pregnancy rates are not completely comparable between the 1989 form of the birth certificate and the 2003 form of the birth certificate with the 2003 revised form

Consider the variables (iii) average of poverty rate in 1980 and 1990 and (iv) average of poverty rate in 2000 and 2010. They measure average poverty rates in the before period (1977-1989) and after period (1999-2011) respectively. A similar reasoning applies for median household income and education. The exact dates chosen are based on data availability, e.g., the 1990 census measured income in 1989, education began being measured by the American Community Survey after 2010 etc.

Smoking is only available from 1989 onwards. There is a high degree of correlation in county smoking rates over time. The correlation between 1989-1995 and 1997-2003 county smoking rates is 0.896

Note that we are matching on some post-treatment (post increase in mountaintop mining) variables, e.g., Average (Mean) of median household income in 1999 and median household income in 2011. Matching on such variables could cause bias if the treatment affects the variables. On the other hand, failing to match on such variables could cause bias if trends between the treated and control groups are only parallel conditional on such variables. For example, trends may not be parallel without matching on Average (Mean) of median household income in 1999 and median household income in 2011 if one group experiences an economic downturn compared to the other that is not related to mountaintop mining. Rosenbaum (1984) discusses this tradeoff. We judge that these variables are largely unaffected by our treatment and match on them to reduce possible bias due to confounding.

<u>Additional variables that we will control for in individual births</u>

(ix) Mother's race (white, black, other)
(x) Maternal age (19 or less, 20-24, 25-29, 30-34, 35-39, 40+)
(xi) Infant sex
(xii) Whether the birth was a multiple birth (twins or more)

*Match*

We pair matched treated counties to control counties. We use the DiPs package in R to do the matching.

|                             | Treated Mean | Control Match Mean | Control All Mean |
|-----------------------------|-------------:|-------------------:|-----------------:|
| median.income.1979.1989     | 18514.957    | 17417.674          | 23192.801        |
| median.income.1999.2011     | 27166.609    | 26616.804          | 39241.436        |
| poverty.1980.1990           | 26.907       | 27.980             | 17.448           |
| poverty.2000.2010           | 26.061       | 26.413             | 16.206           |
| percent.white               | 97.330       | 97.409             | 88.768           |
| education.1980.1990         | 11.395       | 11.354             | 11.818           |
| education.2000.20122016     | 12.191       | 12.176             | 12.721           |
| maternal.smoking.rate       | 0.330        | 0.333              | 0.224            |

---

generally leading to higher smoking rates (Curtin and Matthews, 2016). For this reason, we only consider the smoking rate among pregnant mothers from 1989-2003.

```
                           Control Match SMD Control All SMD
median.income.1979.1989              0.217         -0.926
median.income.1999.2011              0.058         -1.264
poverty.1980.1990                   -0.154          1.360
poverty.2000.2010                   -0.059          1.649
percent.white                       -0.008          0.834
education.1980.1990                  0.095         -0.984
education.2000.20122016              0.032         -1.206
maternal.smoking.rate               -0.034          1.674
```

Before matching, the treated have considerably lower income, higher poverty rate, lower education, higher percent white, less education and more maternal smoking during pregnancy, but after matching, the covariates are relatively balanced.

| Matched Set | Treated | Control |
|---|---|---|
| 1 | Bell County, KY | Knox County, KY |
| 2 | Pike County, KY | Robertson County, KY |
| 3 | Wise County, VA | Taylor County, KY |
| 4 | Boone County, WV | Meigs County, TN |
| 5 | Clay County, WV | Scott County, TN |
| 6 | Fayette County, WV | Summers County, WV |
| 7 | Kanawha County, WV | Daviess County, KY |
| 8 | Lincoln County, WV | Powell County, KY |
| 9 | Logan County, WV | Hawkins County, TN |
| 10 | McDowell County, WV | Lake County, TN |
| 11 | Mingo County, WV | Campbell County, TN |
| 12 | Breathitt County, KY | Johnson County, KY |
| 13 | Nicholas County, WV | Laurel County, KY |
| 14 | Raleigh County, WV | Mercer County, WV |
| 15 | Webster County, WV | Estill County, KY |
| 16 | Wyoming County, WV | Mason County, WV |
| 17 | Floyd County, KY | Lewis County, KY |
| 18 | Harlan County, KY | Clay County, KY |
| 19 | Knott County, KY | Wolfe County, KY |
| 20 | Leslie County, KY | Owsley County, KY |
| 21 | Letcher County, KY | Russell County, KY |
| 22 | Martin County, KY | Union County, TN |
| 23 | Perry County, KY | Menifee County, KY |

Pre-treatment period trend in low birthweight (<2500 grams) for the match:

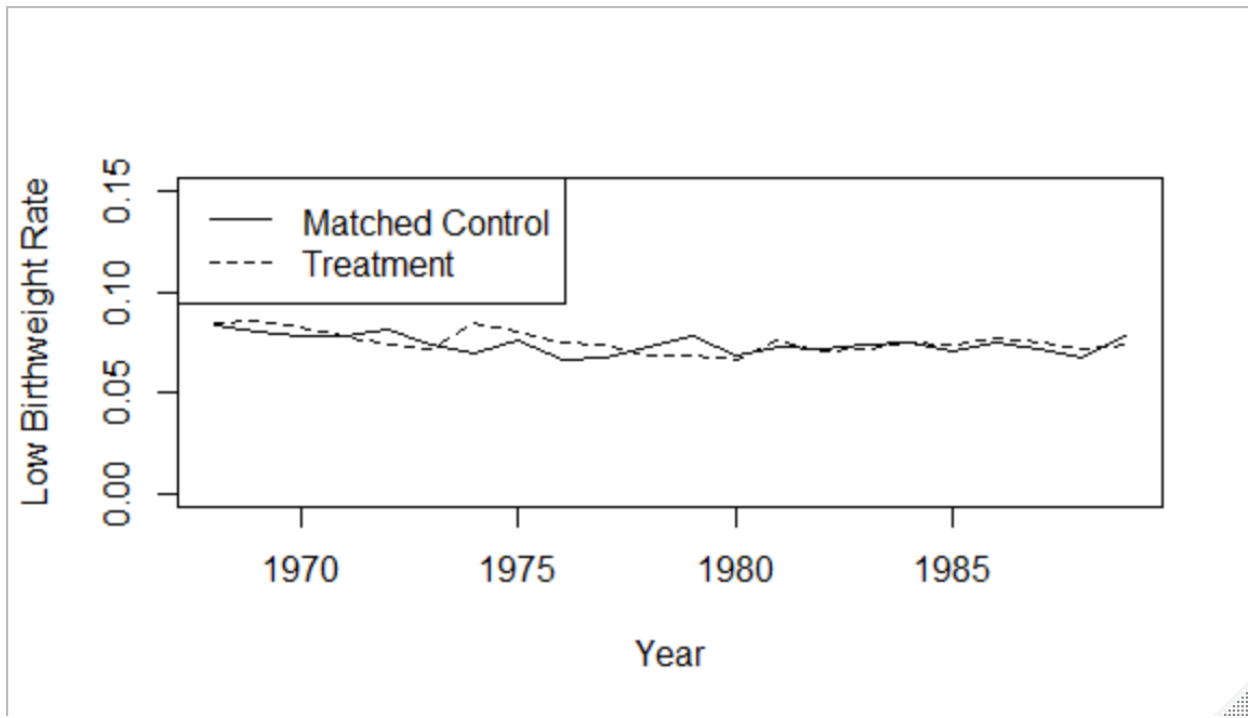

The pre-period low birth weight rates are similar for the treated and matched control counties. This is a paired t-test comparing the rates from 1969-1989 for the treated and matched control counties.

```
> t.test(lowbwrate.treated.county,lowbwrate.matched.control.county,paired=TRU
E)

        Paired t-test

data:  lowbwrate.treated.county and lowbwrate.matched.control.county
t = 0.018729, df = 22, p-value = 0.9852
alternative hypothesis: true difference in means is not equal to 0
95 percent confidence interval:
 -0.004736298  0.004822624
sample estimates:
mean of the differences
           4.316271e-05
```

This is a paired t-test comparing the rates from 1977-1989:

```
> t.test(lowbwrate.treated.county,lowbwrate.matched.control.county,paired=TRU
E)

        Paired t-test

data:  lowbwrate.treated.county and lowbwrate.matched.control.county
t = -0.6676, df = 22, p-value = 0.5113
alternative hypothesis: true difference in means is not equal to 0
95 percent confidence interval:
 -0.007268922  0.003728677
sample estimates:
```

```
mean of the differences
       -0.001770123
```

*Multiple Control Groups*

Our control group for our primary analysis allows counties that border treated counties to be in the control group.

As a secondary control, we consider a match that excluded the border counties. The match is not as good but we will use it in a secondary analysis.

```
> round(check(Xmat,Xmatch,treatment,match2a$data$treatment),3)
                         Treated Mean Control Match Mean Control All Mean
median.income.1979.1989     18514.957          17565.543        23569.206
median.income.1999.2011     27166.609          28062.435        40044.883
poverty.1980.1990              26.907             27.367           16.884
poverty.2000.2010              26.061             24.378           15.663
percent.white                  97.330             96.752           88.054
education.1980.1990            11.395             11.337           11.837
education.2000.20122016        12.191             12.204           12.746
maternal.smoking.rate           0.330              0.313            0.218
                         Control Match SMD Control All SMD
median.income.1979.1989             0.188           -1.003
median.income.1999.2011            -0.094           -1.350
poverty.1980.1990                  -0.069            1.498
poverty.2000.2010                   0.293            1.813
percent.white                       0.055            0.889
education.1980.1990                 0.134           -1.016
education.2000.20122016            -0.031           -1.257
maternal.smoking.rate               0.275            1.807
```

*Outcomes*

The primary outcome will be low birth weight (birth weight <2500g). We focus on birth weight rather than gestational age because in older data going back to the late 1970s and early 1980s, gestational age was less well measured than it is today.
The following will be secondary outcomes

(i) Very low birth weight (<1500g)
(ii) Preterm birth (Gestational age <37 weeks)
(iii) Small for gestational age (<10$^{th}$ percentile of birth weight for gestational age based on the 10$^{th}$ percentile estimates in Talge et al. (2014).

For secondary outcomes (ii) and (iii), gestational age is not available for a substantial portion of Kentucky and Tennessee in 1977-1980 so we will use only 1981-1989 as the pre-period.

*Analytic plan*

For the primary analysis, we will fit a logistic regression model of the outcome on dummy variables for each county, year dummy variables, the individual covariates and a variable for the dose of surface mining in a treated county in the post period. This dose $D_{it}$ will be defined as follows:

0 if observation is in pre-period
0 if observation is in post-period control group
Max(Disturbed Area Percentage in Year minus Average Disturbed Area Percentage in 1985-1989,0) if observation is in post-period treated group

The dose $D_{it}$ can be thought of as reflecting a county $i$'s amount of surface mining at time $t$ relative to its amount in the pre-period. For the pre-period, this dose is 0 since it is relative to the pre-period. For the post-period, the dose is set to 0 for control counties since surface mining was low or non-existent in these counties in both the pre-period and post-period. The following plot show the average surface mining production per square mile in the pre-period years for which this data is available (1983-1989) and the post-period (1999-2011) for the treated and matched control counties. Note that for some matched control counties, the surface mining may have gone down in the post-period relative to the pre-period and so in some sense, the dose could be negative, but making the dose zero will be conservative in terms of finding an effect of surface mining.

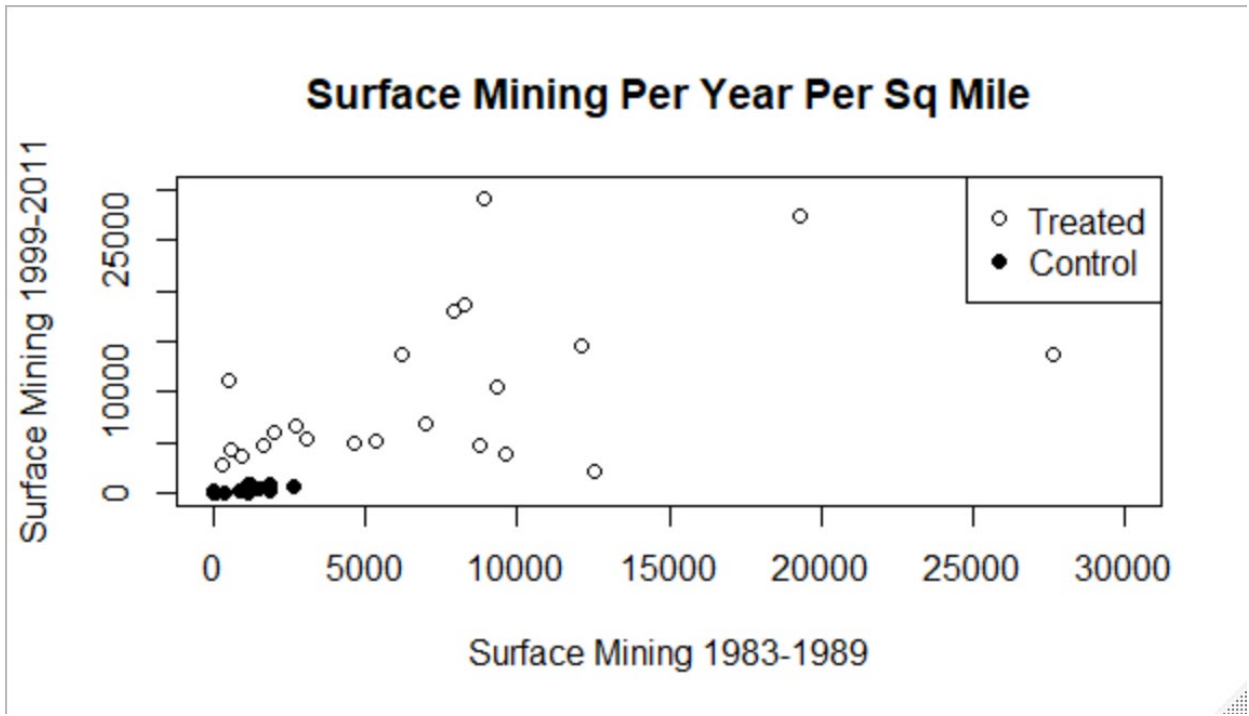

The coefficient on the dose variable will estimate the effect on the log-odds ratio of a 1 percentage point increase in the county disturbed area percentage (assuming disturbed area percentage in 1985-1989 is the same as that for the whole control period of 1977-1989). This follows the analytic approach for a difference-in-difference study of Volpp et al. (2007). The model is

$$\text{logit}[\Pr(Y_{itj} = 1 | X_{itj}, T_{it})] = \alpha_i + \lambda_t + \gamma^T X_{itj} + \chi D_{it} \qquad (0.1)$$

where $Y_{itj}$ is the outcome (say low birthweight) for baby $j$ in county $i$ in year $t$, $X_{itj}$ are covariates, $T_{it}$ is whether county $i$ is treated (high surface mining in the post period) in year $t$, $\alpha_i$ are county fixed effects and $\lambda_t$ are year fixed effects. $\exp(\chi)$ is the parameter of interest – the odds ratio for the outcome being 1 in a treated county with a disturbed area percentage of 1 vs. a control county (or a disturbed area percentage of $x$ vs. a disturbed area percentage $x$-1 for any $x$>0) when the covariates and fixed effects are held constant. To account for the possibility of period specific county effects, we will use cluster robust standard errors, a generalization of the White robust covariance sandwich estimator that allow for group-level correlation (clustering) in addition to heteroscedasticity (White, 1980; Liang and Zeger, 1986).

We will consider two secondary analyses:
(1) We will fit a logistic regression model of the outcome on dummy variables for each county, year dummy variables, the individual covariates and a dummy variable for whether the observation is in a treated county in the post period. The coefficient on this last variable is our estimate of the effect of being in a high surface mining county in the post period as compared to being in a low/no surface mining county.
. The model is

$$\text{logit}[\Pr(Y_{itj} = 1 | X_{itj}, T_{it})] = \alpha_i + \lambda_t + \gamma^T X_{itj} + \beta T_{it} \quad (0.2)$$

where $Y_{itj}$ is the outcome (say low birthweight) for baby $j$ in county $i$ in year $t$, $X_{itj}$ are covariates, $T_{it}$ is whether county $i$ is treated (high surface mining in the post period) in year $t$, $\alpha_i$ are county fixed effects and $\lambda_t$ are year fixed effects. $\exp(\beta)$ is the parameter of interest – the odds ratio for the outcome being 1 in a treated county vs. control county when the covariates and fixed effects are held constant.

(2) We will fit model (1.1) just using the treated counties.

Secondary analyses (1) and (2) are evidence factors in that they are nearly independent analyses that would potentially be affected by different biases (Rosenbaum, 2010, 2015). Analysis (1) would be biased by post-period factors that affect the trends of high surface mining counties vs. low/no surface mining counties differently and analysis (2) would be affected by post-period factors that affects the trends of very high surface mining counties vs. high but not very high surface mining counties differently.

To assess whether trends between the treated and matched control groups were similar, we conducted the below "test of controls" (Volpp et al., 2007), carrying out the analysis on the pre-period with the "treated period" being 1984-1989. The "treated period" coefficient was not significant, providing no evidence of a differential trend between the groups in the control period.

```
> trial.model=lrm(bwt.lt.2500~fullcounty+year+mothrace+mothage.cat+sex+plural
ity+treatment.temp,data=natlfull,subset=(group!="None" & numeric.year<=1989),
x=TRUE,y=TRUE)
```

```
> robcov(trial.model,fullcounty.character)
Frequencies of Missing Values Due to Each Variable
   bwt.lt.2500      fullcounty            year         mothrace     mothage.cat
           330               0               0               41               0
           sex        plurality   treatment.temp
             0               0                0

Logistic Regression Model

 lrm(formula = bwt.lt.2500 ~ fullcounty + year + mothrace + mothage.cat +
     sex + plurality + treatment.temp, data = natlfull, subset = (group !=
     "None" & numeric.year <= 1989), x = TRUE, y = TRUE)

                                  Model Likelihood      Discrimination
Rank Discrim.
                                     Ratio Test             Indexes
    Indexes
 Obs                296343      LR chi2    9737.07      R2         0.080
C        0.626
   FALSE            274970      d.f.            67      g          0.362
Dxy      0.253
   TRUE              21373      Pr(> chi2) <0.0001      gr         1.436
gamma    0.254
 Cluster on  fullcounty.character                       gp         0.033
tau-a    0.034
 Clusters               46                              Brier      0.063

 max |deriv|          6e-12

                                  Coef    S.E.   Wald Z Pr(>|Z|)
Intercept                        0.7716 0.1060   7.28  <0.0001
fullcounty=Boone County, WV     -0.0168 0.0027  -6.33  <0.0001
fullcounty=Breathitt County, KY -0.0421 0.0034 -12.29  <0.0001
fullcounty=Campbell County, TN   0.1092 0.0140   7.81  <0.0001
fullcounty=Clay County, KY       0.1700 0.0136  12.45  <0.0001
fullcounty=Clay County, WV      -0.1107 0.0026 -42.95  <0.0001
fullcounty=Daviess County, KY   -0.1439 0.0141 -10.18  <0.0001
fullcounty=Estill County, KY     0.0738 0.0136   5.43  <0.0001
fullcounty=Fayette County, WV   -0.0403 0.0055  -7.30  <0.0001
fullcounty=Floyd County, KY     -0.0670 0.0028 -24.22  <0.0001
fullcounty=Harlan County, KY    -0.0479 0.0017 -27.50  <0.0001
fullcounty=Hawkins County, TN   -0.0450 0.0140  -3.22   0.0013
fullcounty=Johnson County, KY   -0.1568 0.0136 -11.51  <0.0001
fullcounty=Kanawha County, WV   -0.0941 0.0054 -17.52  <0.0001
fullcounty=Knott County, KY     -0.1148 0.0023 -50.06  <0.0001
fullcounty=Knox County, KY       0.0711 0.0137   5.18  <0.0001
fullcounty=Lake County, TN       0.2981 0.0314   9.48  <0.0001
fullcounty=Laurel County, KY    -0.0877 0.0140  -6.27  <0.0001
fullcounty=Leslie County, KY    -0.0488 0.0026 -18.53  <0.0001
fullcounty=Letcher County, KY   -0.0883 0.0020 -44.90  <0.0001
fullcounty=Lewis County, KY     -0.0220 0.0133  -1.66   0.0976
fullcounty=Lincoln County, WV   -0.0592 0.0030 -19.65  <0.0001
fullcounty=Logan County, WV      0.0781 0.0016  48.33  <0.0001
fullcounty=Martin County, KY    -0.0294 0.0029 -10.20  <0.0001
fullcounty=Mason County, WV     -0.1513 0.0132 -11.44  <0.0001
fullcounty=McDowell County, WV   0.2421 0.0124  19.45  <0.0001
fullcounty=Meigs County, TN      0.0648 0.0136   4.78  <0.0001
fullcounty=Menifee County, KY    0.1010 0.0137   7.38  <0.0001
fullcounty=Mercer County, WV     0.1958 0.0148  13.23  <0.0001
fullcounty=Mingo County, WV      0.0714 0.0011  65.38  <0.0001
fullcounty=Nicholas County, WV  -0.1933 0.0038 -50.58  <0.0001
fullcounty=Owsley County, KY    -0.0640 0.0138  -4.64  <0.0001
```

```
fullcounty=Perry County, KY         0.0508 0.0013  38.43 <0.0001
fullcounty=Pike County, KY         -0.0251 0.0025  -9.92 <0.0001
fullcounty=Powell County, KY       -0.0557 0.0139  -4.00 <0.0001
fullcounty=Raleigh County, WV      -0.0141 0.0077  -1.82  0.0687
fullcounty=Robertson County, KY    -0.1113 0.0140  -7.93 <0.0001
fullcounty=Russell County, KY      -0.1288 0.0146  -8.81 <0.0001
fullcounty=Scott County, TN        -0.0039 0.0135  -0.29  0.7741
fullcounty=Summers County, WV       0.0948 0.0120   7.89 <0.0001
fullcounty=Taylor County, KY       -0.1022 0.0149  -6.88 <0.0001
fullcounty=Union County, TN         0.2653 0.0146  18.22 <0.0001
fullcounty=Webster County, WV      -0.0920 0.0035 -26.62 <0.0001
fullcounty=Wise County, VA          0.0055 0.0023   2.41  0.0161
fullcounty=Wolfe County, KY        -0.0181 0.0131  -1.38  0.1669
fullcounty=Wyoming County, WV      -0.0271 0.0028  -9.84 <0.0001
year=1978                          -0.0179 0.0338  -0.53  0.5960
year=1979                           0.0227 0.0405   0.56  0.5757
year=1980                          -0.0348 0.0301  -1.16  0.2468
year=1981                           0.0853 0.0347   2.46  0.0140
year=1982                           0.0111 0.0329   0.34  0.7357
year=1983                           0.0503 0.0365   1.38  0.1685
year=1984                           0.0471 0.0377   1.25  0.2114
year=1985                           0.0247 0.0456   0.54  0.5884
year=1986                           0.0789 0.0330   2.39  0.0167
year=1987                           0.0237 0.0399   0.60  0.5517
year=1988                          -0.0240 0.0435  -0.55  0.5810
year=1989                           0.0488 0.0342   1.43  0.1533
mothrace=Other                     -0.3251 0.1630  -1.99  0.0461
mothrace=White                     -0.5665 0.0776  -7.30 <0.0001
mothage.cat=25-29                  -0.1405 0.0321  -4.38 <0.0001
mothage.cat=30-34                  -0.0514 0.0437  -1.18  0.2392
mothage.cat=35-39                   0.2128 0.0392   5.43 <0.0001
mothage.cat=40+                     0.3646 0.0800   4.56 <0.0001
mothage.cat=Less Than 20            0.2759 0.0206  13.38 <0.0001
sex=Male                           -0.1764 0.0168 -10.52 <0.0001
plurality=Single Birth             -2.9002 0.0347 -83.47 <0.0001
treatment.temp                      0.0371 0.0316   1.17  0.2409
```

For all of the analyses, we will remove observations with missing data. There is little missing data as shown in the output for the "test of controls" above

*Sensitivity Analysis*

If the effect of treatment (high surface mining) is significant in our primary analysis, we will conduct a sensitivity analysis. We will consider an unmeasured binary confounder $u$ that is present only in the treated counties in the post time period. We will assess how large the effect on $u$ would have to be in order for the effect of treatment to no longer be significant.

*Interpreting the Estimate*

A limitation of our study is that our treatment is defined at the county level but in a treated county, we expect a range of exposures to surface mining for mothers living in the county due their proximal location. Distributions of disturbed mine lands are not uniform across the area of a county.

To understand better what our estimated treatment effect might mean for mothers who are affected by surface mining, we let $W_{itj}$ denote whether mother $j$ in county $i$ at time $t$ is affected by surface mining ($W_{itj} = 0$ if $T_{it} = 0$ and $W_{itj} = 1$ or $0$ if $T_{it} = 1$) and consider the following alternative model to (1.1):

$$\text{logit}[\Pr(Y_{itj} = 1 \mid X_{itj}, W_{itj})] = \alpha_i + \lambda_t + \gamma^T X_{itj} + \tau W_{itj} \tag{0.3}$$

where $\Pr(W_{itj} = 1 \mid D_{it}, T_{it} = 1) = \theta D_{it}$. When $\theta = 1$, it means if the disturbed area fraction increases by 0.01, we expect 1% of the mothers to be affected; when $\theta = 2$, it means if the disturbed area fraction increases by 0.01, we expect 2% of the mothers to be affected, etc.

We cannot estimate $\theta$ from the data so will vary it in a sensitivity analysis. We will consider $\theta$ values of 1,..., 10. Given $\theta$, we will estimate model (1.3) using the EM algorithm (Ibrahim, 1990). The parameter $\exp(\tau)$ is the odds ratio effect of treatment for babies whose mother is affected by surface mining.

To compute standard errors for these interpretability analyses, we will compute a design effect from the main analyses (i.e., cluster robust variance / non-cluster robust variance) and multiply the variances by the design effect.

*Acknowledgements:*
We thank Brenda Eskenazi and Bob Gunier for helpful discussion.